\documentclass[showpacs,10pt,twocolumn,prl]{revtex4-1}
\usepackage{amsmath}
\usepackage{multirow}
\usepackage{amssymb}
\usepackage{graphics}
\usepackage{epsfig}
\usepackage{color}

\begin{document}

\title{Evolution of ultra-flat band in van der Waals kagome semiconductor Pd$_{3}$P$_{2}$(S$_{1-x}$Se$_{x}$)$_{8}$}
\author{Shaohua Yan$^{1,\dag}$, Ben-Chao Gong$^{1,\dag}$, Lin Wang$^{2}$, Jinzhi Wu$^{1}$, Qiangwei Yin$^{1}$, Xinyu Cao$^{3}$, Xiao Zhang$^{3}$, Xiaofeng Liu$^{2,*}$, Zhong-Yi Lu$^{1}$, Kai Liu$^{1,*}$, and Hechang Lei$^{1,*}$}
\affiliation{$^{1}$Department of Physics and Beijing Key Laboratory of Opto-electronic Functional Materials $\&$ Micro-nano Devices, Renmin University of China, Beijing 100872, China\\
$^{2}$School of Materials Science $\&$ Engineering, Zhejiang University, Hangzhou 310027, China\\
$^{3}$State Key Laboratory of Information Photonics and Optical Communications $\&$ School of Science, Beijing University of Posts and Telecommunications, Beijing 100876, China
}
\date{\today}

\begin{abstract}
We investigate the evolutions of structural parameters, optical properties, and electronic structures of van der Waals kagome semiconductor Pd$_{3}$P$_{2}$S$_{8}$ with Se doping. 
When the doping level of Se increases, the bandgaps of Pd$_{3}$P$_{2}$(S$_{1-x}$Se$_{x}$)$_{8}$ single crystals decrease gradually, accompanying with the expanded unit cells.
The first-principles calculations show that there is a flat band (FB) near the Fermi level in bulk Pd$_{3}$P$_{2}$S$_{8}$. This FB mainly originates from the $d_{z^2}$-like orbitals of Pd atoms in the Pd kagome lattice, which has a finite interlayer electron hopping perpendicular to the PdS$_4$ square plane.  
The interlayer hopping can be reinforced with the Se doping, inducing a stronger interlayer coupling via the chalcogen atoms at apical sites, which reduces the bandgap and enhances the cleavage energy. In contrast, the vanishing  interlayer hopping in the two-dimensional limit results in the formation of ultra-FB in the monolayers of these compounds.
The easy exfoliation and the existence of unique ultra-FB near $E_{\rm F}$ make Pd$_{3}$P$_{2}$(S$_{1-x}$Se$_{x}$)$_{8}$ a model system to explore the exotic physics of FB in two-dimensional kagome lattice. 
\end{abstract}

\maketitle

\section{Introduction} 

Flat band (FB) represents one type of unusual band structure with constant energy independent of the crystal momentum (dispersionless band in momentum space) \cite{Leykam,Rhim,LiuZ1}. The zero bandwidth (BW) of ideal FB leads to high density of states (DOS) and vanishing group velocity with infinite effective mass of electrons.
For the quenched kinetic energy and high DOS, the FB systems are thought to have a strong electron-electron correlation effect naturally, which could cause many of exotic many-body phenomena, like ferromagnetism, superconductivity, Wigner crystals etc \cite{Mielke,Tasaki,Miyahara,Ko,Wu,Jaworowski}.
Moreover, when time reversal symmetry is broken and spin-orbital coupling is considered, the FB can become topologically nontrivial with none-zero Chern invariant, leading to quantum Hall state or even high-temperature fractional quantum Hall state \cite{Ohgushi,Tang}. 
In theory, various two-dimensional (2D) lattice models have been proposed to host the FB based on line-graph construction in general \cite{Mielke,Mielke2,Mielke3}, such as dice, Lieb, kagome, and decorated square lattices, as well as honeycomb lattice with multiple orbitals on each site \cite{Sutherland,Lieb,Tasaki,Tasaki2,Mielke2,Mielke3,Tang,Leykam,LiuZ1,Wu,Wu2,LiuZ2}.

The 2D kagome lattice composed of corner-sharing triangles and hexagons of atoms is one of lattice models in which the FB exists \cite{LiuZ1}. 
This FB roots in the destructive phase interference of electron hopping paths and in real space the electronic state is geometrically confined within the single hexagon, forming a compact localized state \cite{Leykam,Rhim}.
Theoretical studies have shown that most of exotic phenomena originating from the FB can appear in kagome lattice \cite{Mielke,Ko,Wu,Ohgushi,Tang}.
Moreover, 2D kagome lattice also exhibits unique topological band structure and strong magnetic frustration effects like topologically nontrivial Dirac band, saddle point with van Hove singularity, quantum spin liquid state \cite{YeL,Yu,WangWS,Kiesel,Balents}. 
Recent experimental studies on insulating or metallic materials with 2D kagome lattice (kagome materials) have confirmed the existence of topological electronic structures including FB \cite{YeL,LiuZ,LiuZH,KangM,LiM}, and revealed various exotic phenomena-fractionalized excitations of spin liquid state, negative FB magnetism, large intrinsic anomalous Hall effect, massive Dirac point with Chern gap, coexisted charge density wave and superconducting states, to name a few \cite{HanTH,Nakatsuji,LiuE,WangQ,YinJX,YinJX2,YinJX3,Ortiz1,Ortiz2,YinQW}.

Despite a lot of fascinating properties of kagome materials, the experimental studies on their intrinsic physical properties in 2D limit especially for FB are still scarce. Due to the finite interlayer couplings and the existence of structural layers other than the kagome one in real materials, the FB will be perturbed, inducing the band dispersion and hybridization with other bands \cite{LiuZ,LiuZH,KangM,LiM}.
Thus, in order to reveal the peculiar properties of FB in kagome lattice, the exploration of kagome materials with vanishing interlayer coupling becomes utterly important.

Recently, semiconducting Pd$_{3}$P$_{2}$S$_{8}$ with Pd kagome lattice is highlighted due to its weak van der Waals (vdW) interlayer interaction and a FB near $E_{\rm F}$ with a very small BW, closely related to the kagome structure of Pd atoms \cite{Park}.
In order to understand the evolution and origin of this unusual FB in Pd$_{3}$P$_{2}$S$_{8}$, in this work, we carry out a systematic study on Pd$_{3}$P$_{2}$(S$_{1-x}$Se$_{x}$)$_{8}$ single crystals with $x$ up to 0.25. It is found that with Se doping the lattice parameters increase gradually but the bandgaps decrease. Theoretical calculations indicate that the narrowing of bandgap and the increased cleavage energy can be ascribed to the enhanced interlayer coupling by Se doping. Further analysis reveals that the ultra-FB in monolayer Pd$_{3}$P$_{2}$Ch$_{8}$ (Ch = S, Se) originates from the Pd $d_{z^2}$-like orbitals of the Pd kagome lattice and the absence of interlayer electron hoping perpendicular to the local PdCh$_4$ square plane.

\section{Methods}

Single crystals of Pd$_{3}$P$_{2}$(S$_{1-x}$Se$_{x}$)$_{8}$ were grown by chemical vapor transport method. Pd powder (99.98 \% purity), P powder (99.999 \% purity), S flakes (99.95 \% purity), and Se powder (99.99 \% purity) in a 3 : 2 : 8(1 - $x$): 8$x$ molar ratio were put into a silicon tube with 80 mg iodine flakes (99.999 \%). The tube was sealed under high vacuum and then placed in a two-zone horizontal tube furnace. The temperatures of two zones were raised slowly to 993 K and 963 K for 2 days and were then held there for another 3 days. After that, the temperatures were decreased slowly down to 673 K and 623 K. Finally, the power of furnace was switched off and the ampoule was cooled down naturally. Crystals with typical size of 3$\times$3$\times$0.5 mm$^{3}$ can be obtained. 

The elemental analysis was performed using the energy dispersive x-ray spectroscopy (EDX). The doping level $x$ mentioned below is the actual composition obtained from the EDX measurements.
Exfoliation of Pd$_{3}$P$_{2}$(S$_{1-x}$Se$_{x}$)$_{8}$ crystals was achieved using mechanical exfoliation with scotch tape and was transferred onto a 300-nm SiO$_{2}$ covered Si substrate. The microscopy images was acquired using a Bruker Edge Dimension atomic force microscope (AFM).
X-ray diffraction (XRD) patterns were measured using a Bruker D8 x-ray machine with Cu $K_{\alpha}$ ($\lambda$ = 1.5418 \AA) radiation. 
Room-temperature optical transmission spectra of the samples were recorded in a Hitachi UH5700 spectrophotometer in the spectral range of 400$-$1100 nm. The exfoliated flakes of samples were fixed onto a stainless steel sample holder with an aperture diameter of 1.0 mm. The absorption coefficient ($\alpha$) was then calculated according to the Beer-Lambert law. 

The density functional theory (DFT) calculations on Pd$_{3}$P$_{2}$Ch$_{8}$ (Ch = S, Se) were performed with the projector augmented wave (PAW) method \cite{D1,D2} as implemented in the Vienna Ab initio Simulation Package (VASP) \cite{D3,D4,D5}. The generalized gradient approximation (GGA) of Perdew-Burke-Ernzerhof (PBE) \cite{D6} type was adopted for the exchange-correlation functional. The kinetic energy cutoff of the plane-wave basis was set to 420 eV. A fully variable-cell relaxation was carried out to obtain the equilibrium lattice parameters. The internal atomic positions were relaxed with the quasi-Newton algorithm until the forces on all atoms were smaller than 0.01 eV/\AA. For the Brillouin zone sampling of the bulk and the monolayer of Pd$_{3}$P$_{2}$Ch$_{8}$, the $8\times 8 \times 8$ and $8\times 8 \times1$ $k$-point meshes were used for the structural optimization calculations, while the $12\times 12 \times 12$ and $12\times 12 \times 1$ $k$-point meshes were used for the self-consistent calculations, respectively. The Gaussian smearing method with a width of 0.05 eV was utilized for the Fermi surface broadening. For the bulk Pd$_{3}$P$_{2}$Ch$_{8}$, the interlayer vdW interaction was described by using the DFT-D3 method of Grimme \cite{D7}. For the monolayer Pd$_{3}$P$_{2}$Ch$_{8}$, a vacuum space of 20 \AA\ thickness was used to eliminate the interaction of periodic images.
The DOS and total energies of the bulk crystals were obtained by using the tetrahedron method with Bl\"{o}chl corrections \cite{D8}.

\section{Results and discussion}


\begin{figure}[tbp]
\centerline{\includegraphics[scale=0.42]{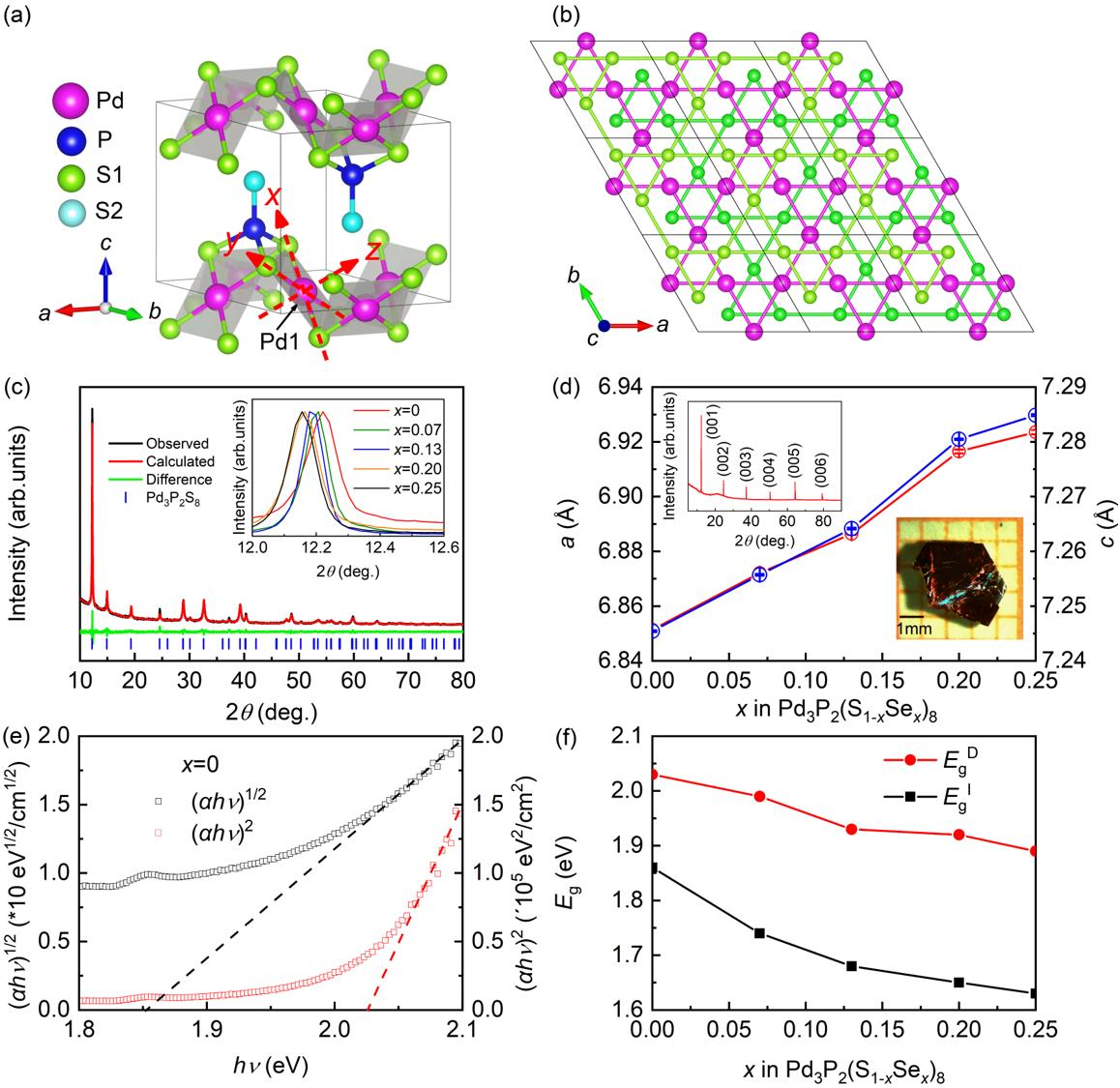}}
\caption{(a) Crystal structure of Pd$_{3}$P$_{2}$S$_{8}$. The local Cartesian coordinates with the origin at the Pd1 atom is defined by the red arrows, where the $z$-axis is perpendicular to the PdS$_4$ square plane. (b) Top view of Pd and S kagome layers. For clarity, the S1 atoms below and above the Pd layer are displayed in different colors. (c) Powder x-ray diffraction (XRD) pattern of Pd$_{3}$P$_{2}$S$_{8}$. Inset: (00$l$) diffraction peaks for Pd$_{3}$P$_{2}$(S$_{1-x}$Se$_{x}$)$_{8}$ with various $x$. (d) The $a$- and $c$-axial lattice parameters $a$ and $c$ as a function of $x$. Inset: (Left) XRD pattern and (Right) optical photo of a Pd$_{3}$P$_{2}$S$_{8}$ single crystal. (e) The plots of $(\alpha h\nu)^{1/2}$ and $(\alpha h\nu)^2$ as a function of $h\nu$ for Pd$_{3}$P$_{2}$S$_{8}$ single crystal. (f) The estimated direct and indirect bandgaps $E_{\rm g}^{\rm D}$ and $E_{\rm g}^{\rm I}$ of Pd$_{3}$P$_{2}$(S$_{1-x}$Se$_{x}$)$_{8}$ single crystals with various $x$.}
\end{figure}

Pd$_{3}$P$_{2}$S$_{8}$ has a layered structure with the stacking Pd-P-S blocking layers along the $c$ axis [Fig. 1(a)]. The key structural ingredient is the perfect Pd kagome lattice in the Pd-P-S layer with the atomic distance of Pd atoms $d_{\rm Pd-Pd}\sim$ 3.418 \AA\ [Fig. 1(b)] \cite{Bither,Simon}.
Below and above the Pd kagome layer, P and S atoms form PS$_{4}$ tetrahedra and each P atom occupies the centre of a tetrahedron.
As shown in Fig. 1(a), because of inequivalent local environment of S atoms, there are two S sites and each of three S atoms at S1 site in PS$_{4}$ tetrahedra is coordinated with two Pd atoms and one P atom with $d_{\rm Pd-S}\sim$ 3.113 \AA\ and $d_{\rm P-S}\sim$ 2.112 \AA, when the fourth S atom at S2 site is coordinated with one P atom only with $d_{\rm P-S}\sim$ 1.904 \AA. 
Thus, the PS$_{4}$ tetrahedron is distorted with the S-P-S angles deviating from 109.47$^{\circ}$ in an ideal tetrahedron to 103.10$^{\circ}$ and 115.27$^{\circ}$ \cite{Bither,Simon}.
More importantly, the vertical P-S bonds in two Pd-P-S layers protrude each other and it leads to the waved Pd-P-S layers and a unique interlocked layered structure of Pd$_{3}$P$_{2}$S$_{8}$, different from conventional vdW materials.
In addition, in each Pd-P-S layer, the S atoms at S1 site also form two distorted kagome lattices below and above the Pd kagome layer with two different values of $d_{\rm S-S}$ ($\sim$ 3.176 \AA\ and 3.665 \AA) [Fig. 1(b)].
On the other hand, each Pd atom is surrounded with four S atoms from two PS$_{4}$ tetrahedra, forming a PdS$_{4}$ tetragon which tilts away from the $ab$ plane. 
In order to keep electric neutrality, the oxidation state of Pd, P, and S in Pd$_{3}$P$_{2}$S$_{8}$ is +2, +5, and -2, respectively. The Pd$^{2+}$ ion has an electronic configuration of [Kr]4$d^{8}$, which usually prefers to have a square-planar crystal field with low spin state. This is consistent with the structural feature and diamagnetism of Pd$_{3}$P$_{2}$S$_{8}$ \cite{Park}.

Figure 1(c) shows the powder XRD pattern and Rietveld fit of ground Pd$_{3}$P$_{2}$S$_{8}$ single crystals. 
The fitted $a$- and $c$-axial lattice parameters $a$ and $c$ are 6.851(3) \AA\ and 7.246(5) \AA, close to the previous results \cite{Bither,Simon}.
Powder XRD patterns of all Se-doped Pd$_{3}$P$_{2}$S$_{8}$ samples can also be fitted very well by using the crystal structure of Pd$_{3}$P$_{2}$S$_{8}$ with the trigonal symmetry (space group $P\text{-}3m1$, No. 164). 
Both fitted $a$ and $c$ increase with increasing Se content monotonically [Fig. 1(d)], which can be ascribed to the larger ionic radius of Se$^{2-}$ than S$^{2-}$. The nearly linear trend of lattice expansion follows the Vegard's law. Such increases of lattice parameters are also partially reflected by the shift of peak position of (001) to lower angle gradually with Se doping [inset of Fig. 1(c)].
It is noted that when $x>$ 0.25, Se can not be doped into sample further even the nominal $x$ reaches 0.5 and Pd(S, Se)$_{2}$ crystals start to grow. This suggests that the solubility limit of Se may be $x=$ 0.25. 
The upper left inset of Fig. 1(d) shows the XRD pattern of a Pd$_{3}$P$_{2}$S$_{8}$ single crystal. All of peaks can be indexed by the indices of (00$l$) lattice planes, indicating that the crystal surface is parallel to the $ab$ plane and perpendicular to the $c$ axis.
The morphology of Pd$_{3}$P$_{2}$S$_{8}$ single crystal is a thick plate with hexagonal shape [lower right inset of Fig. 1(d)], consistent with the layered structure and the trigonal symmetry of Pd$_{3}$P$_{2}$S$_{8}$.

Optical transmittance spectra are measured to determine the bandgaps of Pd$_{3}$P$_{2}$Ch$_{8}$ using the Tauc plot method \cite{Tauc}. As shown in Fig. 1(e), the direct bandgap $E_{\rm g}^{\rm D}$ and indirect one $E_{\rm g}^{\rm I}$ can be estimated from the extrapolation of linear region of $(\alpha h\nu)^2$ and $(\alpha h\nu)^{1/2}$ as a function of photon energy $h\nu$ to zero. Here, $\alpha$ is the absorption coefficient calculated from optical transmission spectra of the samples according to the Beer-Lambert law.
The obtained values of $E_{\rm g}^{\rm D}$ and $E_{\rm g}^{\rm I}$ for Pd$_{3}$P$_{2}$S$_{8}$ are 2.03 and 1.86 eV, in agreement with the values reported previously ($E_{\rm g}^{\rm D}=$ 2.08 eV and $E_{\rm g}^{\rm I}=$ 1.85 eV) \cite{Park} and close to the theoretical one ($E_{\rm g}^{\rm I}=$ 1.46 eV) (Table I).
Both $E_{\rm g}^{\rm D}$ and $E_{\rm g}^{\rm I}$ of Pd$_{3}$P$_{2}$(S$_{1-x}$Se$_{x}$)$_{8}$ decrease monotonically with increasing $x$ [Fig. 1(f)]. This is in line with the color changes of single crystals from red orange to dark red with Se doping.

\begin{table}[]
\caption{\label{tab:I} The calculated lattice constants (in unit of \AA), indirect bandgap $E_{\rm g}^{\rm I}$  (in unit of eV) and bandwidth (BW, in unit of meV) of Pd$_{3}$P$_{2}$S$_{8}$, Pd$_{3}$P$_{2}$(S$_{0.75}$Se$_{0.25}$)$_{8}$, and Pd$_{3}$P$_{2}$Se$_{8}$ along with the experimental lattice constants.}
\begin{tabular}{|l|l|l|l|l|l|l|l|}
\hline
\multicolumn{2}{|l|}{\multirow{2}{*}{}}                                        & \multicolumn{2}{l|}{Expt.}   & \multicolumn{4}{l|}{Cal.}                                                            \\ \cline{3-8} 
\multicolumn{2}{|l|}{}                                                                     & $a$              & $c$                 & $a$          & $c$                       &$E_{\rm g}^{\rm I}$    & BW              \\ \hline
\multirow{2}{*}{Pd$_{3}$P$_{2}$S$_{8}$}        & bulk                 & 6.842           & 7.247             & 6.918       & 7.314                   & 1.46                   &  348             \\ \cline{2-8} 
                                                                               & monolayer      & ---                 & ---                  & 7.115       & ---                         & 1.85                   &  68                \\ \hline
\multirow{2}{*}{P-1}                                            & bulk                 & 6.930           & 7.283             & 6.960       & 7.491                   & 1.09                   &325                  \\ \cline{2-8} 
                                                                               & monolayer      & ---                 & ---                  & 7.138        & ---                        & 1.69                   & 69                      \\ \hline
\multirow{2}{*}{P-2}                                            & bulk                 & 6.930           & 7.283             & 6.991       & 7.354                   & 1.36                   &403                       \\ \cline{2-8} 
                                                                               & monolayer      & ---                 & ---                  & 7.192        & ---                        & 1.72                   &   69                    \\ \hline
\multirow{2}{*}{Pd$_{3}$P$_{2}$Se$_{8}$}     & bulk                  & ---                 & ---                  & 7.184        & 7.666                  & 0.89                   & 478                \\ \cline{2-8} 
                                                                              & monolayer       & ---                 & ---                  & 7.385        & ---                        & 1.47                   & 172                       \\ \hline
\end{tabular}
\end{table}

Theoretical calculations confirm the increased lattice parameters and decreased bandgaps of Pd$_{3}$P$_{2}$(S$_{1-x}$Se$_{x}$)$_{8}$ with Se doping (Table I). 
To mimic the structure of Pd$_{3}$P$_{2}$(S$_{0.75}$Se$_{0.25}$)$_{8}$, two typical structural configurations namely P-1 with Se atoms at S1 site (P-1 phase) and P-2 with Se atoms at S2 site (P-2 phase) are considered (Fig. S1 in SM) \cite{SM}.
The calculated total energy of the P-1 phase are 363.2 meV/f.u. lower than that of the P-2 phase for Pd$_{3}$P$_{2}$(S$_{0.75}$Se$_{0.25}$)$_{8}$. It indicates that Se atoms prefer to dope into the S1 site, which may result from the longer bond length and weaker binding strength of the P-S1 bond than those of the P-S2 bond. 
The calculated lattice parameters of hypothetical Pd$_{3}$P$_{2}$Se$_{8}$ are larger than those of Pd$_{3}$P$_{2}$S$_{8}$ due to the larger ionic radius of Se$^{2-}$ than S$^{2-}$. 
Meanwhile, the theoretical value of $E_{\rm g}^{\rm I}$ of Pd$_{3}$P$_{2}$Se$_{8}$ is also smaller than that of Pd$_{3}$P$_{2}$S$_{8}$ (Table I), consistent with the experimental trend [Fig. 1(d)].

\begin{figure}
\centerline{\includegraphics[scale=0.32]{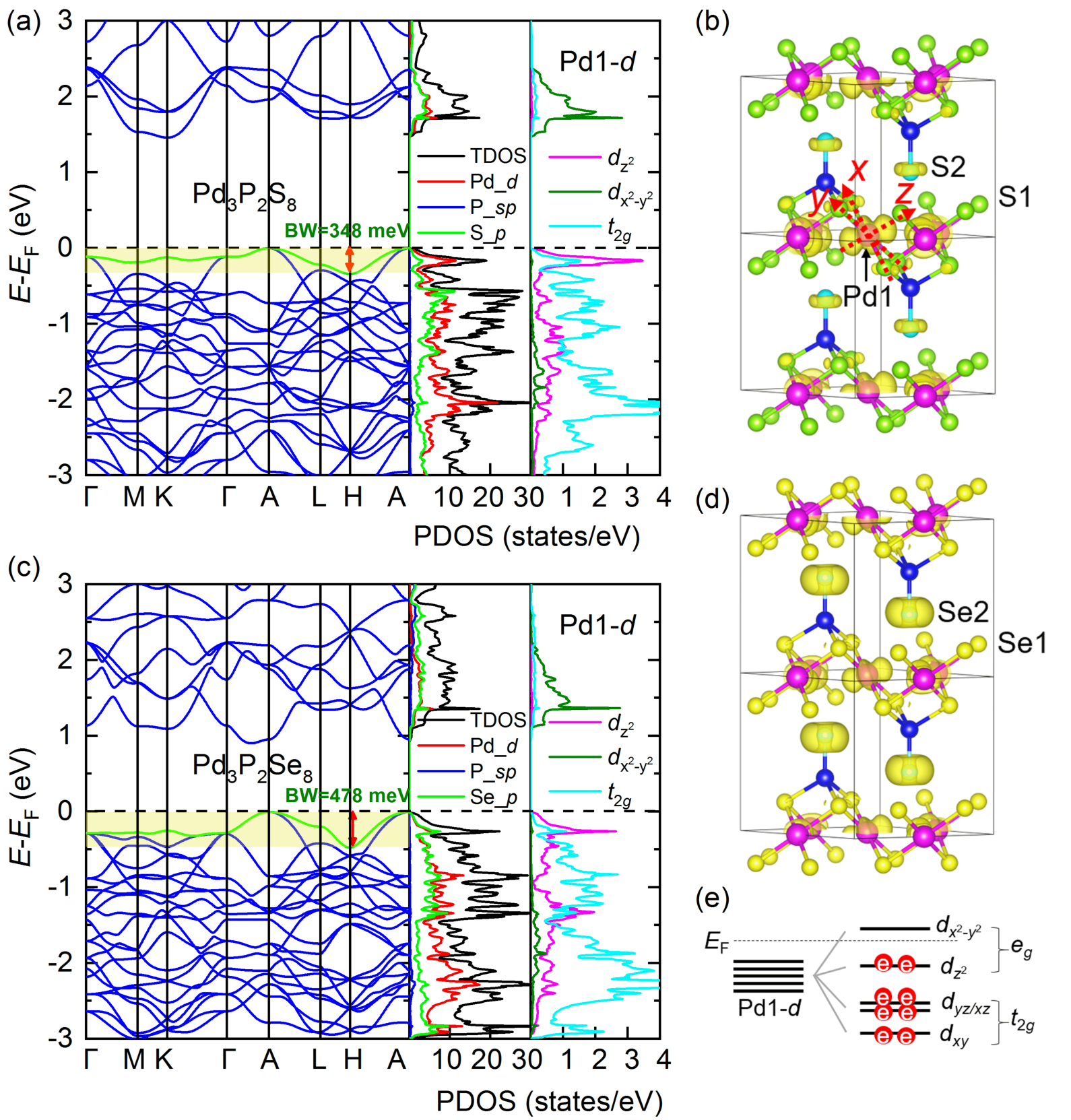}} \vspace*{-0.3cm}
\caption{Band structures, total density of states (TDOS), and PDOS of Pd-$d$, P-$sp$, and S(Se)-$p$ orbitals for bulk (a) Pd$_{3}$P$_{2}$S$_{8}$ and (c) Pd$_{3}$P$_{2}$Se$_{8}$. The BWs are marked in green. (b) and (d) The ICDs of flat bands for bulk Pd$_{3}$P$_{2}$S$_{8}$ and Pd$_{3}$P$_{2}$Se$_{8}$, respectively. The isosurface of charge densities is set to 0.003 e/\AA$^3$. (e) Schematic diagram of crystal field splitting for the Pd1 atom.}
\end{figure}

In order to understand the evolutions of bandgap and FB of Pd$_{3}$P$_{2}$(S$_{1-x}$Se$_{x}$)$_{8}$ with Se doping, we first study the electronic structures of bulk Pd$_{3}$P$_{2}$S$_{8}$. 
Even with the three-dimensional structural features and the contributions of multiple orbitals from Pd, P, and S atoms, some of the unique features of the 2D kagome lattice of Pd still can be observed in Pd$_{3}$P$_{2}$S$_{8}$, including the FB centered just below the Fermi level $E_{\rm F}$ and the Dirac point at the $K$ point of Brillouin zone [Fig. 2(a)].
From the partial density of states (PDOS), it can be seen that the valance bands near the $E_{\rm F}$ mainly consist of Pd-$d$ and S-$p$ orbitals, while the P orbitals are far away from the $E_{\rm F}$. The common peaks of Pd-$d$ and S-$p$ PDOSs indicate the presence of $p$-$d$ hybridizations.
These $p$-$d$ hybridizations vied for influence and dominance below -0.28 eV, while for the FBs, it mainly depends on Pd-$d$ orbitals. 
In comparison with other kagome materials, Pd$_{3}$P$_{2}$S$_{8}$ has a FB with rather small (BW = 348 meV) [Table I]. To explore its origin, we calculate the integrated charge densities (ICDs) of FBs for bulk Pd$_{3}$P$_{2}$S$_{8}$ with the local Cartesian coordinates, whose axes are along the Pd-S bonds and perpendicular to the PdS$_4$ square plane [Fig. 1(a) and Fig. 2(b)]. 
Interestingly, the FB exhibits a nearly single-orbital (Pd-$d_{z^2}$-like orbital) behavior with the direction perpendicular to the PdS$_4$ square plane. 
Actually, this can be well explained by the splitting of energy level of Pd$^{2+}$ ion in the local square-planar crystal field [Fig. 2(e)], similar to the case of NdNiO$_2$ \cite{NdNiO}. The topmost occupied orbital of Pd$^{2+}$ ion is $d_{z^2}$ and this is consistent with the PDOS shown in Fig. 2(a). 
In addition, the bands just below the FB are mainly consist of Pd-$t_{2g}$ (nondegenerate $d_{xy}$ and doubly degenerate $d_{yz}$/$d_{xz}$) orbitals and S-$p$ orbitals [Fig. 2(a)]. This $p$-$d$ hybridization results in the large dispersion of these bands.
In contrast, because of the large distance between the Pd atom and the S2 atoms in the neighboring layers, the electron hopping perpendicular to the PdS$_4$ square plane is limited and there is only small amount of charges around S2 atoms contributing to the FB [Fig. 2(b)], giving rise to the small BW of FB. In fact, similar phenomenon is also observed at the surface of the kagome metal FeSn by destructing electronic hopping along the vertical direction \cite{FeSn}. 

Considering Se doping, the main features of electronic structure of bulk Pd$_{3}$P$_{2}$(S$_{1-x}$Se$_{x}$)$_{8}$ ($x=$ 0.25 and 1) [Figs. 2(c) and S2 in SM] are similar to that of Pd$_{3}$P$_{2}$S$_{8}$ \cite{SM}. 
But the weight of Se-$p$ orbitals below the $E_{\rm F}$ in Pd$_{3}$P$_{2}$Se$_{8}$ is larger than that of S-$p$ orbitals in Pd$_{3}$P$_{2}$S$_{8}$ [Figs. 2(a) and 2(c)], suggesting the stronger $p$-$d$ hybridizations in the former. 
Meanwhile, the BW of FB (478 meV) [Table I] and the ICDs around Se2 in Pd$_{3}$P$_{2}$Se$_{8}$ are also larger than those of Pd$_{3}$P$_{2}$S$_{8}$ [Figs. 2(b) and 2(d)].
Furthermore, the FB of P-2 phase exhibits a larger dispersion along the $\Gamma$-A direction than that of P-1 phase [Table I and Fig. S2 in SM] \cite{SM}. This large dispersion can also partially explain the reduced $E_{\rm g}^{\rm I}$ with Se doping, especially at S2 site [Fig. 1(f) and Table I] .  
Above results clearly indicate that with increasing Se content the increase of Pd-Ch (Ch = S, Se) $p$-$d$ hybridization, especially when the Se atoms locate at apical sites, has a significant influence on interlayer coupling, BW of FB and bandgap, which seems to be a disadvantage to realize the ultra-FB of Pd kagome lattice. 

\begin{figure}
\centerline{\includegraphics[scale=0.17]{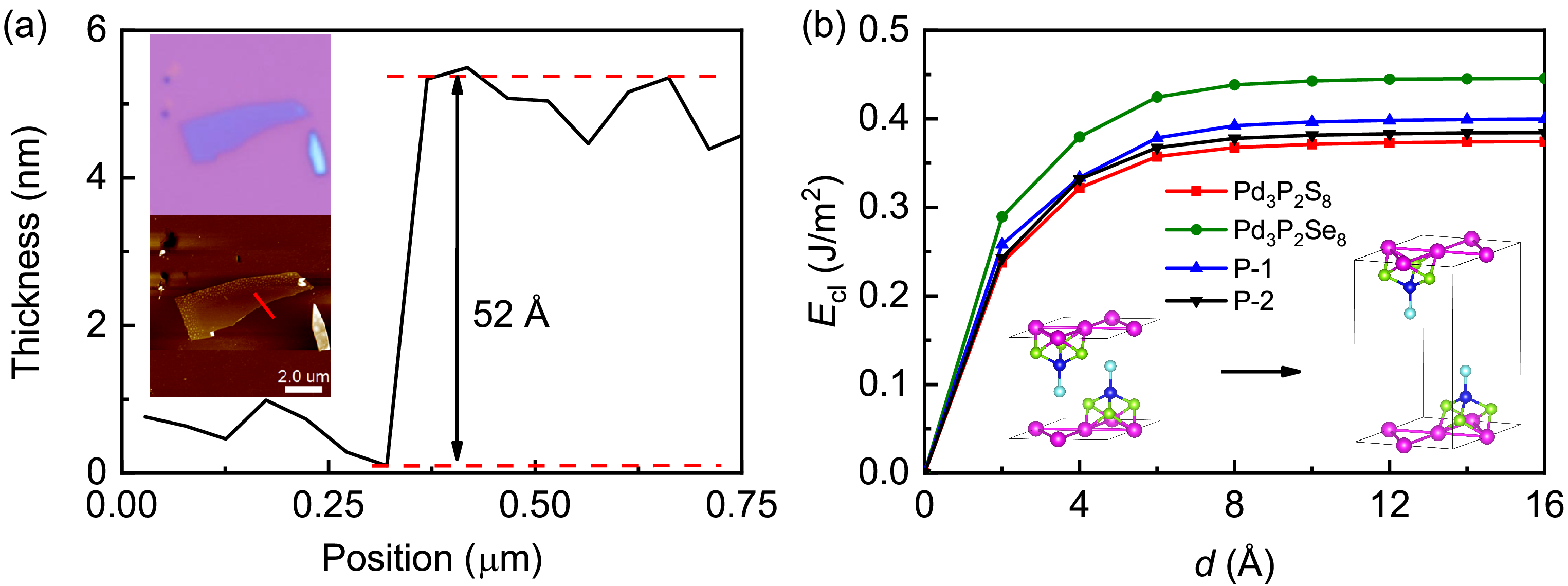}} \vspace*{-0.3cm}
\caption{(a) Height profile across sample edge along the direction shown by the arrow in the lower inset. Upper and lower insets show the optical and AFM images of a cleaved thin flake of Pd$_{3}$P$_{2}$S$_{8}$ crystal. The bar in lower inset is 2 $\mu$m. (b) Calculated $E_{\rm cl}$ as a function of $d$ for Pd$_{3}$P$_{2}$S$_{8}$, Pd$_{3}$P$_{2}$Se$_{8}$, as well as Pd$_{3}$P$_{2}$(S$_{0.75}$Se$_{0.25}$)$_{8}$ with P-1 and P-2 configurations.}
\end{figure}

In order to reduce the interlayer coupling, a natural idea is to exfoliate the layered compounds of Pd$_{3}$P$_{2}$Ch$_{8}$ down to monolayers. 
Previous experiment reported that Pd$_{3}$P$_{2}$S$_{8}$ can be easily cleaved down to few layers or even monolayer \cite{Park}. 
For the whole series of Pd$_{3}$P$_{2}$Ch$_{8}$ crystals, they are also easy to exfoliate to few layers like Pd$_{3}$P$_{2}$S$_{8}$.
Here, we take Pd$_{3}$P$_{2}$S$_{8}$ as an example for cleavage. The step height across an edge of the cleaved sample [red line in the lower inset of Fig. 3(a)] is 52 \AA\ [Fig. 3(a)], which is about 7 unit cells of Pd$_{3}$P$_{2}$S$_{8}$. 
Moreover, the rather large area of exfoliated thin flake with the same color ($\sim$ 7$\times$3 $\mu$m$^{2}$) suggests a uniform thickness through the whole sample [upper inset of Fig. 3(a)].
It is noted that the thin flakes of Pd$_{3}$P$_{2}$Ch$_{8}$ are stable in air without color change for several days.
To reveal the evolution of interlayer coupling strength with Se doping further, the cleavage energies $E_{\rm cl}$ are calculated by increasing interlayer distance $d$ [inset of Fig. 3(b)].
The $E_{\rm cl}$ increases quickly with increasing $d$ and then converges to the saturation value of 0.37 J/m$^2$ (23 meV/\AA$^2$) when $d$ is lager than $\sim$ 8 \AA. 
With Se doping, the $E_{\rm cl}$ increases gradually and reaches 0.45 J/m$^2$ (28 meV/\AA$^2$) for hypothetical Pd$_{3}$P$_{2}$Se$_{8}$. 
In addition, the P-1 phase has a slightly larger $E_{\rm cl}$ = 0.40 J/m$^2$ than the P-2 phase ($E_{\rm cl}\sim$ 0.38 J/m$^2$). 
These results confirm the significant influence of chalcogen atoms at apical site on the strength of interlayer coupling.  
Although the $E_{\rm cl}$ increases with Se doping, they are still comparable with those values for most of the well-known vdW materials, like graphite and MoS$_{2}$ ($E_{\rm cl}$ = 13 - 21 meV/\AA$^{2}$) \cite{Bjorkman}. Thus Pd$_{3}$P$_{2}$Ch$_{8}$ can be classified to the group of easily exfoliable materials ($E_{\rm cl}<$ 35 meV/\AA$^{2}$) \cite{Mounet}, even they have the wavy structure of Pd-P-Ch layers.

\begin{figure}
\centerline{\includegraphics[scale=0.23]{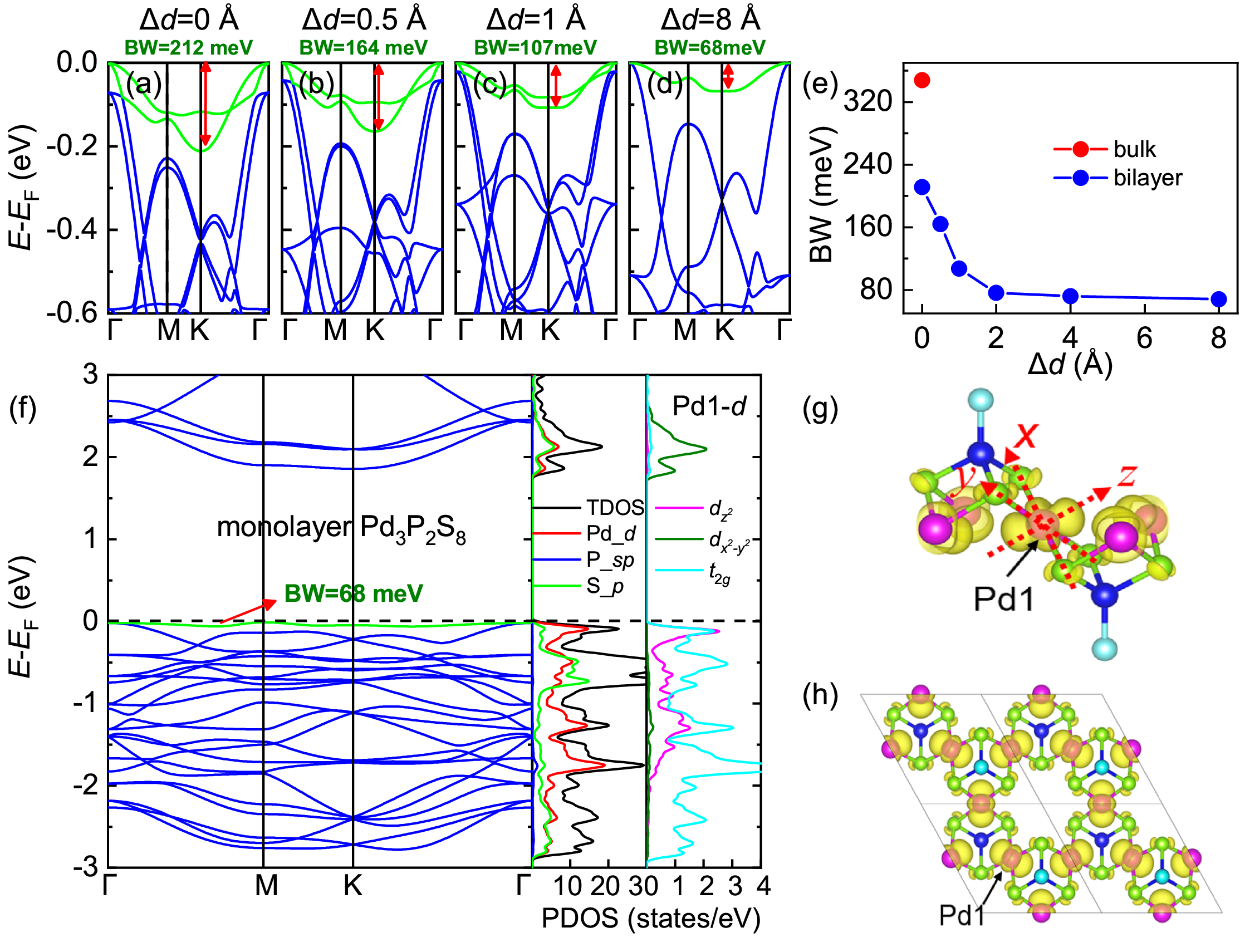}} \vspace*{-0.3cm}
\caption{(a)-(d) Band structures of the bilayer Pd$_{3}$P$_{2}$S$_{8}$ with increasing $\Delta d$. The green bands mark the two FBs. (e) The BW of two FBs as a function of $\Delta d$, as well as the value of bulk Pd$_{3}$P$_{2}$S$_{8}$ (red dot). (f) Band structure, TDOS, and PDOS of Pd-$d$, P-$sp$, and S-$p$ orbitals for monolayer Pd$_{3}$P$_{2}$S$_{8}$. (g) Side view and (h) top view of the ICDs of FB for the monolayer Pd$_{3}$P$_{2}$S$_{8}$.}
\end{figure}

The feasible cleavage of Pd$_{3}$P$_{2}$Ch$_{8}$ down to atomically thin films or even monolayer alters the interlayer coupling and also has a significant influence on its electronic structure.
To further clarify the evolution of FBs with the strength of interlayer interaction, we consider the continuous evolution of band structure in bilayers Pd$_{3}$P$_{2}$S$_{8}$ with the variation of interlayer spacing $\Delta d=(d-d_{\rm Bulk})$, where $d_{\rm Bulk}$ is the interlayer distance in bulk materials [Fig. 4(a)]. 
When the $\Delta d=$ 0, the strong interlayer interaction leads to the coupling of two sets of FBs derived from two different layers with relatively large BW (212 meV).
It is noted that this value is close to the average value of bulk and monolayer Pd$_{3}$P$_{2}$S$_{8}$ (Table I) possibly because in the bilayer case almost half of interlayer couplings are absent when compared to the bulk one.
With increasing $\Delta d$, the weakened coupling of two sets of FBs narrows the BW [Figs. 4(b) and 4(c)].
Finally when the $\Delta d$ is large enough ($\Delta d$ = 8 \AA), the interlayer coupling becomes negligible and the two sets of FBs are almost degenerated [Fig. 4(d)].
Correspondingly, the BW of two FBs decreases by more than threefold [Fig. 4(e)].
If exfoliated to monolayer, the Pd-S2 distance becomes infinite and the charge around S2 atoms vanishes [Figs. 4(g) and 4(h)], suggesting the lack of interlayer electron hopping along the out-of-plane direction. As a result, the much narrow BW of FB ($\sim$ 68 meV) appears in the monolayer of Pd$_{3}$P$_{2}$S$_{8}$ [Figs. 4(f)]. 
On the other hand, the BW of FB for monolayer Pd$_{3}$P$_{2}$Se$_{8}$ is slightly larger than that of monolayer Pd$_{3}$P$_{2}$S$_{8}$ and the ICDs near Se1 is also larger than that of S1 in Pd$_{3}$P$_{2}$S$_{8}$ [Figs. S3 in SM] \cite{SM}.
It suggests that the increased intralayer coupling between Pd and Se1 due to stronger $p$-$d$ hybridization are also unfavorable for the flatness of FB.
We have also calculated the band structures of Pd$_{3}$P$_{2}$Ch$_{8}$ with spin-orbit coupling (Fig. S4 in SM)\cite{SM}. The spin-orbit coupling has minor effect on the band structures.  In addtion, the magnetic calculations suggest that the ground states of these materials are nonmagnetic, which is consistent with the low spin state of Pd$^{2+}$ ions ($S=$ 0). 

\section{Conclusion}

In summary, we study the doping effects of Se on the vdW kagome semiconductor Pd$_{3}$P$_{2}$S$_{8}$ single crystals. It is found that with Se doping the lattice parameters increase but the bandgaps $E_{g}$ decrease monotonically, which be related to the increased dispersion of FB near $E_{\rm F}$ due to the enhanced interlayer coupling due to Se doping, especially when Se atoms occupy the apical sites of PCh$_{4}$ tetrahedra.
In contrast, once such interlayer coupling is eliminated in monolayer Pd$_{3}$P$_{2}$Ch$_{8}$, an unusual ultra-FB of Pd kagome lattice with narrow BW can be formed.
This ultra-FB not only roots in the kagome structure of Pd$_{3}$P$_{2}$Ch$_{8}$ but also is closely related to the unique square planar crystal field of PdCh$_{4}$ with Pd$^{2+}$ ion in $4d^{8}$ configuration.
The latter one isolates the $d_{z^2}$ band of from other $d$ bands of Pd$^{2+}$ ion when the interlayer electron hopping perpendicular to the PdCh$_{4}$ square plane is suppressed in the 2D limit.
Thus, Pd$_{3}$P$_{2}$Ch$_{8}$ provides a unique platform to study the physics of FB in kagome lattice.

\section{Acknowledgements}

This work was supported by National Key R\&D Program of China (Grants No. 2018YFE0202600, No. 2017YFA0302903, and No. 2019YFA0308603), Beijing Natural Science Foundation (Grant No. Z200005), National Natural Science Foundation of China (Grants No. 12174443, and No. 11934020), the Fundamental Research Funds for the Central Universities and Research Funds of Renmin University of China (RUC) (Grants No. 18XNLG14, No. 19XNLG13, and No. 19XNLG17), and Beijing National Laboratory for Condensed Matter Physics. Computational resources were provided by the Physical Laboratory of High-Performance Computing at Renmin University of China.

$^{\dag}$ S.H.Y and B.C.G. contributed equally to this work.

$\ast$ Corresponding authors: xfliu@zju.edu.cn (X. F. Liu); kliu@ruc.edu.cn (K. Liu); hlei@ruc.edu.cn (H. C. Lei).

\end{document}